\documentclass{mem}
\usepackage{natbib}\usepackage{txfonts}\usepackage{balance}
\usepackage{graphicx}
\usepackage[a4paper]{hyperref}
\usepackage{placeins}
\idline{00}{00}

\begin{document}
\def\teff{$T\rm_{eff }$}
\def\kms{$\mathrm {km s}^{-1}$}

\title{
Carbon stars in local group dwarf galaxies:
}

   \subtitle{C and O abundances\thanks{
Based on observations collected at the European Southern Observatory, Paranal, Chile (ESO Programme 70.D-0414 \& 072.D-0501)
} }

\author{
R. \,Wahlin\inst{1}
\and K. \,Eriksson\inst{1}  
\and B. \,Gustafsson\inst{1}
\and K. H. \,Hinkle\inst{2} 
\and D. L. \,Lambert\inst{3} 
\and N. \,Ryde\inst{1}      
\and B. \,Westerlund\inst{1}
          }

  \offprints{R. Wahlin}

\institute{
Department of Astronomy and Space Physics, Uppsala University,
Box 515,
SE-751 20 Uppsala,
Sweden
\email{rurik.wahlin@astro.uu.se}
\and
National Optical Astronomy Observatory, PO Box 26732, Tucson, AZ 85726, USA
\and
Department of Astronomy, University of Texas, Austin, TX 78712, USA
}

\authorrunning{Wahlin }

\titlerunning{Carbon stars in local group dwarf galaxies}

\abstract{
We present abundances of carbon and oxygen as well as 
abundance ratios $\mathrm{^{12}C/^{13}C}$ for a sample of carbon stars in 
the LMC, SMC, Carina, Sculptor and Fornax dwarf galaxies.
The overall metallicities in these dwarf galaxies are lower than in the
galactic disc. The observations cover most of the AGB and we discuss the
abundance patterns in different regions along the AGB. 
The abundances are determined from infrared spectra obtained with the
ISAAC spectrometer on VLT (R=1500) and the Phoenix Spectrometer on Gemini
South (R=50000). The synthetic spectra used in the analysis were computed
with MARCS model atmospheres. We find that the oxygen abundance is 
decreasing with decreasing overall metallicity of the system while the 
C/O ratio at a given evolutionary phase is increasing with decreasing 
oxygen abundance. 

}
\maketitle{}

\section{Introduction}

The AGB phase is the last stage of the evolution of low and intermediate mass stars powered by interior nuclear burning. Chemical elements produced in the stellar interior are mixed to the surface and ejected into the interstellar medium. In this process carbon is also brought to the surface and may transform the star into a carbon star, where the carbon abundance exceeds the oxygen abundance in the stellar atmosphere. During the evolution on the AGB the luminosity increases and AGB stars are among the most luminous stars.
Evolutionary models of AGB stars depend on several free parameters which are adjusted to fit certain criteria like the LMC carbon star luminosity function. An alternative way to constrain the models could be to compare the predicted abundances with those of carbon stars at different luminosities.

A sample of 30 galactic carbon stars was analysed by \citet{lam86} and they found that the carbon excess $\log(\epsilon_C-\epsilon_O)$ was typically small and most carbon stars had a fairly high $^{12}\mathrm{C}/^{13}\mathrm{C}$ ratio. The oxygen abundances were in most cases solar or sub-solar. With the new solar abundances \citep{asp06} as reference the oxygen abundances are typically solar.
One problem with the galactic carbon stars is that it is difficult to determine the distance to them. This introduces an uncertainty in the surface gravities of the stars but also makes the interpretation of the results difficult. In Fig. \ref{figHR} we have plotted our programme stars in the HR diagram showing that it is possible to divide the sample into well defined subgroups along the AGB. Such a division would be difficult to make for the galactic stars, without introducing a contamination of brighter and fainter stars.

In order to study how the atmospheric abundances of carbon stars depend on overall metallicity and luminosity we have selected carbon stars from the local group dwarf galaxies
\object{LMC}, \object{SMC}, \object{Sculptor dSph}, \object{Carina dSph}, \object{Fornax dSph}
with metallicities in the range $-2\lesssim\mathrm{[Fe/H]}\lesssim-0.5$, and well known distances.

\section{Observations}
\begin{figure}[t!]
\resizebox{\hsize}{!}{\includegraphics[angle=270,clip=true]{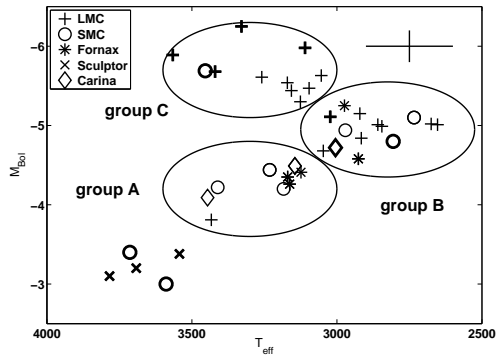}}
\caption{\footnotesize
HR diagram with the programme stars, carbon stars from local group dwarf galaxies.
A clear sequence is seen and the sample can easily be divided into a number of groups with small variation of luminosity. The bold face symbols represent J stars.
}
\label{figHR}
\end{figure}

The programme stars were selected from the literature \citep{fro82,mou82,ric83,azz85,wes87,wes91,wes95}
with the aim of getting a large spread in colour and magnitude, see Fig. \ref{figHR}. The coolest most luminous stars were not included since we cannot yet determine the abundances from spectra of such stars.

\begin{figure*}[t!]
\resizebox{0.5\hsize}{!}{\includegraphics[angle=270,clip=true]{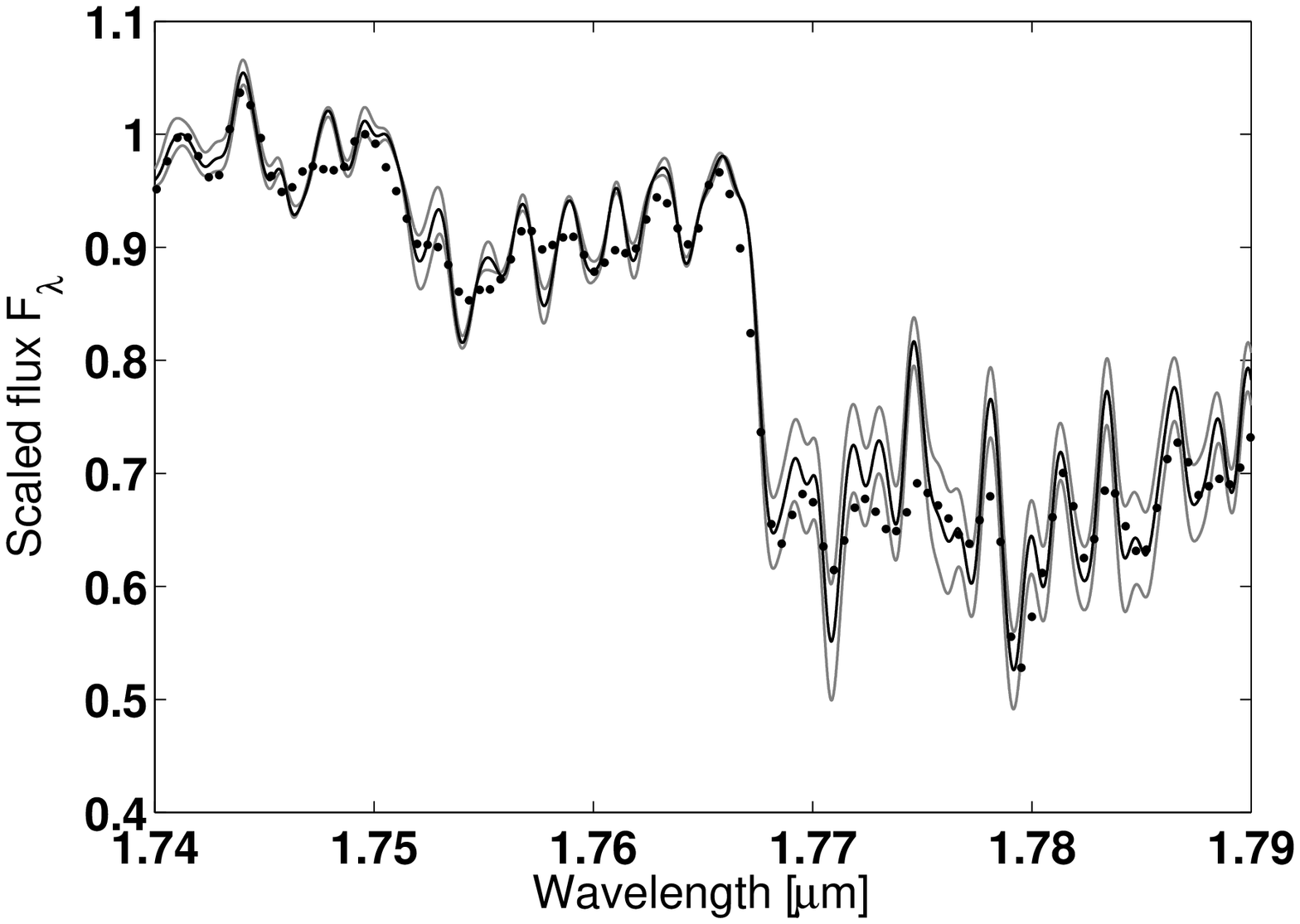}}
\resizebox{0.5\hsize}{!}{\includegraphics[angle=270,clip=true]{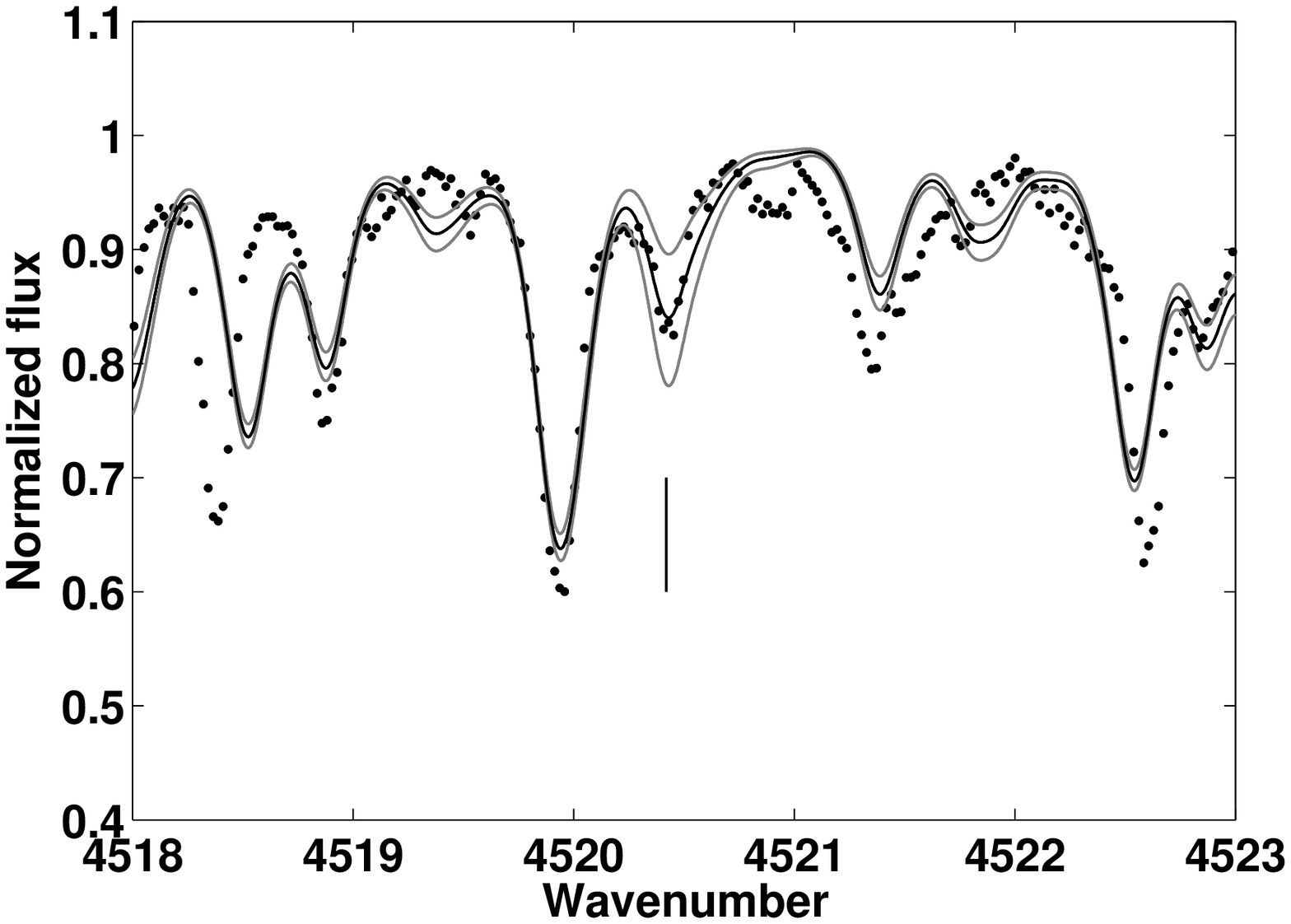}}
\caption{\footnotesize
The left panel shows the ISAAC H band spectrum of the LMC star WORC 47. The spectrum is dominated by the C$_2$ Ballik-Ramsay (0-0) band at $\mathrm{\lambda\gtrsim1.77\mu m}$ with the weaker satelite at  $\mathrm{\lambda\gtrsim1.75\mu m}$. The right panel shows the K band Phoenix high resolution spectrum of the same star. A C$_2$ Phillips line from the (0-2) transition is marked in the plot. The remaining lines are from CN, some of which have uncertain wavenumbers. The lines represents synthetic spectra and the dots observed spectra. The gray lines represents synthetic spectra computed with $\log(\epsilon_C-\epsilon_O)=\pm0.3$dex
}
\label{figISAAC}
\label{figPhoenix}
\end{figure*}
\begin{figure*}[t!]
\resizebox{\hsize}{!}{\includegraphics[angle=270,clip=true]{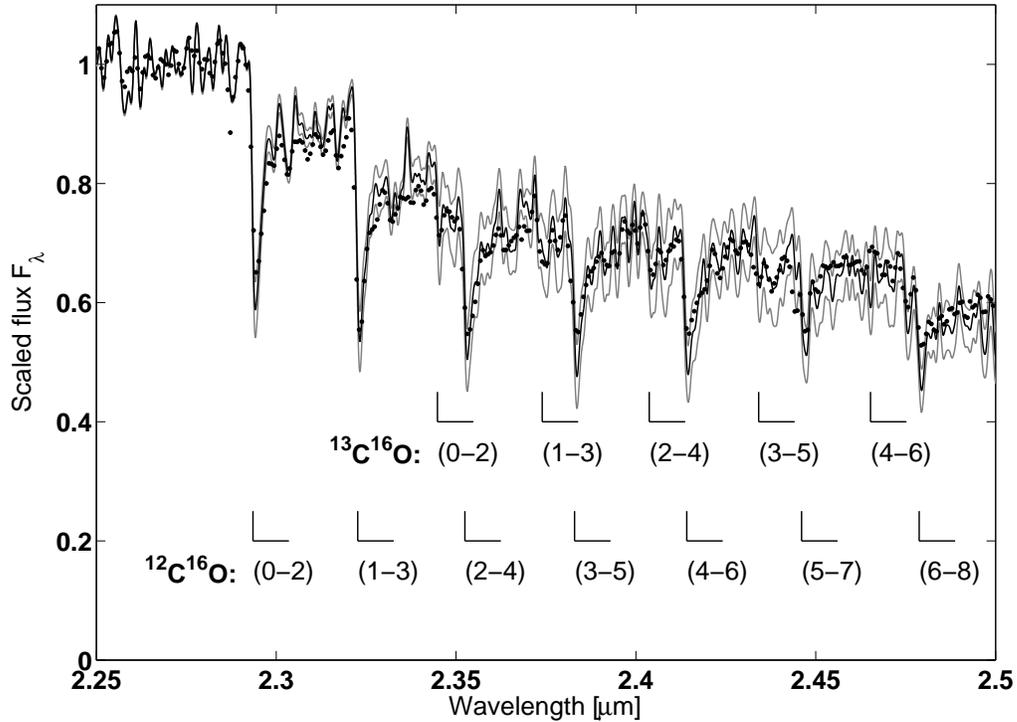}}
\caption{\footnotesize
The ISAAC K band spectrum of the LMC star WORC 47. The spectrum is dominated by the CO first overtone vibration rotation bands. The much weaker CN and C$_2$ bands form a false continuum visible at $\mathrm{\lambda\lesssim2.29\mu m}$. 
The lines represent synthetic spectra and the dots the observed spectrum. The gray lines represent synthetic spectra computed with $\log(\epsilon_O)=\pm0.45$dex.
}
\label{figISAAC-K}
\end{figure*}

Our sample of more than 50 carbon stars were observed in the entire H and K bands using the low resolution mode of the ISAAC spectrograph on VLT UT1 ($R\sim1500$) with S/N$\sim100$.

Since all spectral lines are blended in the low resolution spectra the abundance analysis is uncertain. To verify that our analysis gives reasonable results we selected five stars that were observed with the Phoenix spectrograph on Gemini South giving a resolution of $R\sim50,000$ and S/N$\sim100$. These spectra cover a smaller fraction of the H and K bands but allow individual spectral lines to be measured, see Fig. \ref{figPhoenix}.

The verification sample was selected from the LMC to give an as large spread in stellar parameters, abundances  and isotopic ratios as possible with a small sample. The selection was based on low resolution  criteria from the literature \citep{wes91, ric79}.

The observations were carried out in service mode and data reduction was performed using standard techniques. The telluric features were removed using standard stars observed at the same airmass as the science targets.

\section{Abundance analysis}
The effective temperatures and bolometric magnitudes were determined from photometry 
\citep{cut03,
cio00,
DENIS,wes91,wes95}
using the calibration by
\citet{ber01,ber02}.
The surface gravities were determined assuming a stellar mass of 2$\mathrm{M}_\odot$ following the suggestion by \citet{lam86} and distance modulii from the literature.\footnote{A detailed discussion on the choice of distances will be given in Wahlin et al. (2006 in prep.).
}
\FloatBarrier 
The photometry were dereddened using the dust model presented 
in \citet[Table 4]{dra03}, and $E(B-V)$ from \citet{mat98}.
The overall metallicities were chosen to be the same as the mean metallicity of the host galaxy, using the metallicities from \citet{gre03}.

The model atmospheres and synthetic spectra were computed using the MARCS code originally presented by \citet{gus75}.
The models are hydrostatic, spherically symmetric and computed under the assumption of LTE.
A general description of the grid will be made by Gustafsson et al. (2006, in prep.), and more details given concerning the C stars by J{\o}rgensen et al. (2006, in prep.).

The line lists for the diatomic molecules used in the calculation of synthetic spectra of the abundance criteria were from \citet{goo94} for CO and \citet{wp04} for  C$_2$. 
The synthetic spectra were convolved with a Gaussian profile matching the resolution of the observations. In the high resolution case the spectra were also convolved with a radial tangential profile to account for the macroturbulence. Synthetic spectra computed with different abundances were compared with the observed spectra and the best fits were judged by eye. In this way the carbon excess was determined using molecular features from the C$_2$ Ballik-Ramsay and Phillips electronic transitions in the H band, see Fig. \ref{figISAAC}. The oxygen abundance and $^{12}\mathrm{C}/^{13}\mathrm{C}$ ratio were determined using the CO first overtone vibration rotation transitions in the K band, see Fig. \ref{figISAAC-K}.

\begin{figure}[t!]
\resizebox{\hsize}{!}{\includegraphics[height=10cm,width=10cm,angle=0,clip=true]{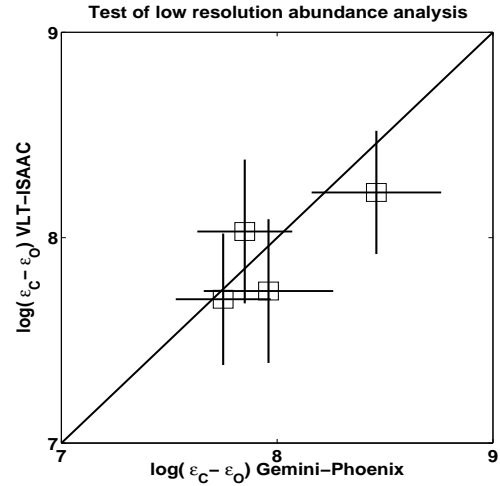}}
\caption{
\footnotesize
The carbon excess of the verification sample as determined in the low resolution analysis plotted against the carbon excess as determined in the high resolution analysis. The error bars do not include uncertainties that would affect both analyses in the same way.
}
\label{figPvsI}
\end{figure}

The stars of the verification sample were also analysed independently with the high resolution spectra. The carbon abundance was determined using lines from the C$_2$ Phillips transition, see Fig. \ref{figPhoenix}, and the oxygen abundance was determined using the second overtone CO vibration rotation lines. The abundances determined from the two different samples agree within the uncertainties, see Fig. \ref{figPvsI}, indicating that the low resolution analysis can be trusted.

\section{Results and Discussion}

\begin{figure}[t!]
\resizebox{\hsize}{!}{\includegraphics[angle=270,clip=true]{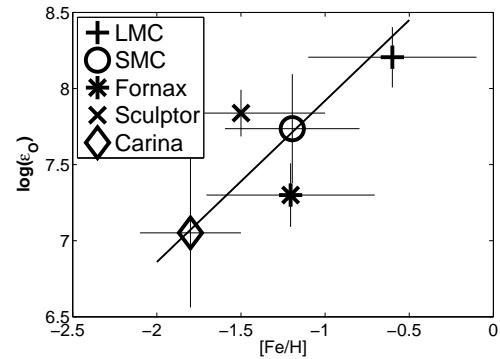}}
\caption{
\footnotesize
The average oxygen abundance within each dwarf galaxy is plotted against the average metallicity in the same galaxy. The error bars on the oxygen abundance indicate the standard deviation of the mean while the error bars on [Fe/H] is the spread in [Fe/H] reported by \citet{gre03}. The solid line is the least squares fit weighted against the uncertainty.
}
\label{figOvsFe}
\end{figure}

We find [O/Fe]=0.25 from the least squares fit in Fig. \ref{figOvsFe}, which is within the uncertainties compared to the abundances determined from H II regions in local group dwarf galaxies \citep{mat98,wes90,duf84}.

\begin{figure*}[t!]
\resizebox{\hsize}{!}{\includegraphics[width=18cm,height=29.7cm,angle=270,clip=true]{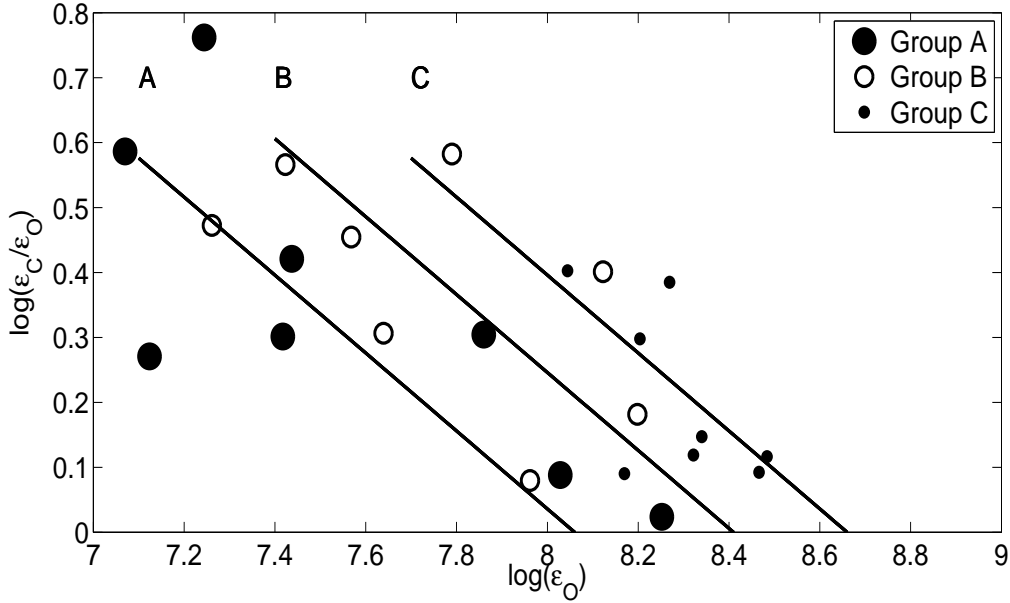}}
\caption{\footnotesize
The evolution of abundances. In this plot we have not included the J stars, since they do not follow the same general behaviour as the remaining sample. The different symbols indicate the stars belong to the groups A, B \& C defined in Fig. \ref{figHR}.
The lines marked with the same names are a fit to the C star data, excluding the J stars, of the form  
$\log{\frac{\epsilon_\mathrm{C}}{\epsilon_\mathrm{O}}}
=
\alpha\cdot\log{\epsilon_\mathrm{O}}+\beta\cdot\mathrm{M_{bol}}+\gamma$, plotted with $\mathrm{M_{bol}}$ fixed at the the mean bolometric magnitude of the stars in the corresponding group.
}
\label{figCO}
\end{figure*}

When the stars are divided into groups along the AGB, excluding the J stars, we find that within each luminosity group the C/O ratio of the normal C stars increases when the oxygen abundance decreases. This is in fair quantitative agreement with the metallicity dependence of the C/O yields calculated by \citet{izz04}
We also find that the C/O ratio at a given oxygen abundance increases with increasing luminosity, see Fig. \ref{figCO}. With this subset of the complete sample we find 

$\log{\frac{\epsilon_\mathrm{C}}{\epsilon_\mathrm{O}}}
=
\alpha\cdot\log{\epsilon_\mathrm{O}}+\beta\cdot\mathrm{M_{bol}}+\gamma$, 

\noindent
where $\alpha=-0.6\pm0.3$, $\beta=-0.3\pm0.1$ and $\gamma=3.55\pm0.15$
Since each $\mathrm{M_{bol}}$ correspond to a range of evolutionary stages and corresponding stellar masses we expect a scatter around the mean of this relation, due to the unknown masses.

Most J stars are found in two distinct regions on the HR diagram where no normal C stars are found, see Fig. \ref{figHR}, one  low luminosity group and one high luminosity group. The low luminosity group shows a large scatter in C/O ratio and no correlation with the oxygen abundance. The high luminosity group show low C/O ratios, especially when compared to the normal C stars with lower luminosity.
The low luminosity J stars may be explained by binary evolution since they probably are too faint to be on the TP-AGB. The high luminosity J stars may be intermediate mass AGB stars that are affected by hot bottom burning (HBB) that would convert the carbon to nitrogen and at the same time increase the $\mathrm{^{12}C/^{13}C}$ ratio.

\citet{lam86} suggested that the small excess of carbon of their sample might be a selection effect due to the immense massloss expected in the stars with large carbon excess. This massloss would enshroud the star and thus remove it from samples of optically bright stars. An alternative possibility could be that the massloss quickly ends the AGB phase by removing the outer layers completely. If this is the reason why the stars are not found and we assume that the massloss is independent of metallicity we would expect the stars in our sample to show a similar behaviour. Since the wind is believed to be driven by radiation pressure on dust grains and the dust is believed to mainly consist of amorphous carbon, this assumption seems reasonable. 

If the amount of carbon brought to the surface by thermal pulses is independent of metallicity we would expect the carbon excess to be much larger in low metallicity stars since their oxygen abundance is so much smaller. The low metallicity carbon stars analysed here show similar enhancements in $\log{(\epsilon_\mathrm{C}-\epsilon_\mathrm{O})}$ as the galactic carbon stars, thus supporting the view of a massloss that is independent of metallicity. There may of course be other explanations for these similarities.

The C/O ratios of our LMC programme stars are typically lower than the C/O ratios of LMC planetary nebulae reported by \citet{sta05}. The difference is similar to the difference found by \citet{lam86} in the comparison between galactic carbon stars and planetary nebulae, even though the LMC stars show a higher median C/O ratio than the galactic sample.

These results will give strict constraints on evolutionary models for AGB evolution.

\FloatBarrier 

\begin{acknowledgements}
Financial support from the
Swedish National Space Board and the
Swedish Research Council is gratefully acknowledged.

This paper is based on observations obtained with the Phoenix infrared spectrograph, developed and operated by the National Optical Astronomy Observatory,
%
%
at the Gemini Observatory, which is operated by the Association of Universities for Research in Astronomy, Inc., under a cooperative agreement with the NSF on behalf of the Gemini partnership: the National Science Foundation (United States), the Particle Physics and Astronomy Research Council (United Kingdom), the National Research Council (Canada), CONICYT (Chile), the Australian Research Council (Australia), CNPq (Brazil) and CONICET (Argentina).
The data were obtained in programs GS-2003A-Q-30 \& GS-2003B-Q-43

\end{acknowledgements}

\bibliographystyle{aa}
\bibliography{wahlin} 

\end{document}